\documentclass[aps,twocolumn,showpacs]{revtex4}

\usepackage{graphicx}
\usepackage{dcolumn}
\usepackage{bm}
\usepackage{amsmath}
\usepackage{hyperref}
\usepackage{hyperref}

\newcommand{\bs}{{\bf S}}

\newcommand{\bss}{({\bf S}_i\cdot{{\bf S}_j})}

\begin{document}
\title{ Topological Quantum Phase Transition in an $S=2$ Spin Chain}
\author{Jiadong Zang$^1$, Hong-Chen Jiang$^2$$^,$$^3$, Zheng-Yu Weng$^3$, Shou-Cheng Zhang$^4$}
\affiliation{$^1$Department of Physics, Fudan University, Shanghai
200433, China\\
$^2$Microsoft Research, Station Q, University of California, Santa
Barbara, CA 93106\\
$^3$Center for Advanced Study, Tsinghua University, Beijing 10084,
China\\
$^4$Department of Physics, Stanford University, Stanford, California
94305, USA}

\date{\today}
\begin{abstract}
We construct a model Hamiltonian for $S=2$ spin chain, where a
variable parameter $\alpha$ is introduced. The edge spin is $S=1$
for $\alpha=0$, and $S=3/2$ for $\alpha=1$. Due to the topological
distinction of the edge states, these two phases must be separated
by one or several topological quantum phase transitions. We've
studied the quantum phase transition by DMRG calculation, and
proposed a phase diagram for this model.
\end{abstract}


\maketitle

\section{Introduction}
Recently, investigations of topological phases and phase transitions
has attracted great attention in condensed matter
physics\cite{Qi2010a}. The quantum Hall state\cite{Klitzing1980a} is
the first example of a topological state of quantum matter, with a
fully gap ground state in the bulk, and gapless excitations at the
edge. The chiral edge state is a holographic mirror of bulk
topology\cite{Thouless1982a}. In the recently discovered time
reversal invariant topological
insulators\cite{Bernevig2006a,Kane2005a}, helical edge states are
confined at the edge by the bulk energy gap, and states with
opposite spins counter-propagate. In the case of the quantum spin
Hall state realized in HgTe quantum wells, the topologically trivial
and non-trivial states are separated by a topological quantum phase
transition, tunable by the thickness of the quantum well.

Quantum spin chain is another example where topological quantum
phase transition is found. The low energy dynamics of 1D large-spin
Heisenberg antiferromagnet can be described as O(3) nonlinear sigma
model\cite{Haldane1988a}.  Half-integer-spin chains are generally
gapless, while integer spin chains are gapped; they are described by
the O(3) nonlinear sigma model with and without the topological
term. This distinction bears strong similarity to the topological
insulators, which is distinct from the conventional insulators by
the presence of a topological term\cite{Qi2008a}.

In this paper, we investigate a spin-2 chain model with two
topologically distinct, transactionally invariant ground states. One
model has edge state $S=1$, while the other model has edge spin
$S=3/2$. The Berry's phase associated with the edge spins differ by
$\pi$. The square of the time reversal operator $T$ gives $T^2=1$
for the first case, whereas it gives $T^2=-1$ in the second case.
Due to this topological difference, the two ground states must be
separated by one or several topological quantum phase transitions
where the spin gap closes.

This paper is constructed as follows. In the next section, we will
review two exact solvable quantum spin model in one dimension. Some
materials can be systematically found in a brilliant work by Tu
et.al\cite{Tu2008}.  Afterwards, our new model Hamiltonian is
presented according to the topological argument. Corresponding
numerical results are shown in the third section, where phase
diagram is discussed as well. Conclusions are drawn in the end.

\section{Model Hamiltonian}
\begin{figure}[tbp]
\includegraphics[width=5cm]{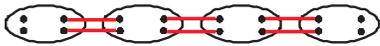} \caption{The
sketch of spin-2 AKLT wavefunction. The circles represent sites, and
each spin-2 is decomposed into four spin-1/2 solid dots. Red lines
connecting the dots stand for singlet states.}\label{Fig:Spin2AKLT}
\end{figure}

Our starting point is an integer spin model introduced by Affleck,
Kennedy, Lieb, and Tasaki, namely the AKLT
model\cite{Affleck1987,Affleck1988}. It is proved that the AKLT
model has a unique infinite volume ground state, with an exponential
decay spin-spin correlation\cite{Affleck1988}. In agreement with the
Haldane conjecture\cite{Haldane1983}, the excitation gap of AKLT
model is nonzero. The ground state of AKLT model can be written down
exactly in terms of Schwinger bosons:
\begin{equation}\label{eq:AKLT_ground_state}
|\Psi ^{AKLT}\rangle=\prod_{\langle ij\rangle}(a_{i}^{\dagger
}b_{j}^{\dagger }-b_{i}^{\dagger }a_{j}^{\dagger })^{S}|0\rangle,
\end{equation}%
where $S$ stands for site spin, and $a^\dagger_i$, $b^\dagger_i$ are
creation operators of Schwinger bosons at the $i$th site. This
ground state can be rephrased in a pictorial form. The spin-2 AKLT
ground state can be schematically shown in Fig\ref{Fig:Spin2AKLT},
where the circles stand for sites on the chain. Each spin-2 can be
decomposed into totally symmetric combinations of four spin-1/2
states, and each state is represented by a solid dot in the figure.
Two pairs of neighboring dots form singlet states, shown in red
bonds. These bonds are usually called \textit{valence bonds}, and in
this sense, AKLT ground state is usually referred as \textit{Valence
Bond State}. It is proved rigorously that this ground state is
unique under this periodic boundary condition. Due to the symmetric
intra-site coupling and anti-symmetric inter-site coupling, the
parent Hamiltonian of this ground state is given by
\begin{equation}  \label{AKLTHam}
H^{AKLT}=\sum_{\langle
ij\rangle}K_3P_3(\mathbf{S}_i,\mathbf{S}_j)+K_4P_4(\mathbf{S}_i,\mathbf{S}_j),\quad
K_3,K_4>0
\end{equation}
where $P_3$  and $P_4$ are the projection operators onto the spin-3
and spin-4 subspaces respectively. The positive coefficients ensure
the state in Eq.(\ref{eq:AKLT_ground_state}) to be the corresponding
groundstate.

In 1998, Scalapino, Zhang, and Hanke (SZH) introduced a SO(5)
symmetric spin model\cite{Scalapino1988} with an exact valence bond
ground state. The original motivation for the model is to illustrate
the SO(5) theory of high Tc superconductivity, which unifies the
antiferromagnetic (AF) and the d-wave superconducting (SC)
phases\cite{Zhang1997,Demler2004}. However, it was soon found later
that the SO(5) symmetry can also be interpreted as an enhanced spin
rotational symmetry\cite{Wu2003a}. The SZH model contains five
quantum states at each site, forming the vector representation of
the SO(5) group. However, these five states can also be interpreted
as the quantum states of the spin $S=2$ of the SO(3) spin chain. SZH
presented an exact ground state wave function expressed as a matrix
product state of the Dirac $\Gamma$ matrices, and showed that the
edge states of the SZH model are 4 fold degenerate at each edge.
Interpreted as an $S=2$ chain language, the edge spin contains
$S=3/2$ spin quantum numbers. Following this work, more valence bond
states with higher symmetry groups have been
constructed\cite{Kluemper1991,Affleck1991,Schuricht2008,Tu2008,Arovas2009}.

The basic idea of the SZH model is the following. The tensor product
of two SO(5) spinor can be decomposed into SO(5) singlet,
antisymmetric tensor, and symmetric traceless tensor, namely,
\begin{equation}
5\times5=1+10+14.
\end{equation}
In analogy with the spin-1 AKLT model, the largest subspace is
projected out, leading to the desired SZH model:
\begin{equation}\label{SZHSO5}
H=J\sum_{\langle xy\rangle}P_{14}(xy),\quad J>0.
\end{equation}
Due to the Clifford algebra of the five $\Gamma$-matrics:
$\Gamma^a\Gamma^b=2\delta^{ab}+2i\Gamma^{ab}$, no symmetric
traceless components are involved in the product of two
$\Gamma$-matrices. As a consequence,
\begin{equation}  \label{SZHwave}
|\Psi^{SZH}\rangle=\sum_{m_1,\ldots
m_N}Tr(\Gamma^{m_1}\Gamma^{m_2}\ldots\Gamma^{m_N})|m_1m_2\ldots
m_N\rangle,
\end{equation}
is the ground state of the above SZH model, where $m$ is a vector
label of the SO(5) group, which can also be interpreted as the
$m_i=-2,-1,0,1,2$ quantum numbers of $S=2$ spin chain.

Due to the relationships between the SO(5) and the SO(3) groups,
there exists a natural deformation of the SZH model to an SU(2)
spin-2 SZH model. The required map from SO(5) group onto SU(2) group
with spin-2 is given by:
\begin{eqnarray}
10(SO(5))&=&3(SO(3))\oplus7(SO(3)) \\
14(SO(5))&=&5(SO(3))\oplus9(SO(3))
\end{eqnarray}
And therefore the SZH Hamiltonian deforms to;
\begin{equation}  \label{SZHHam}
H^{SZH}=\sum_{\langle
ij\rangle}J_2P_2(\mathbf{S}_i,\mathbf{S}_j)+J_4P_4(\mathbf{S}_i,\mathbf{S}_j),\quad
J_2,J_4>0.
\end{equation}
The ground state is unchanged up to an SO(5) rotation.

To see if the SZH state is gapped or not, we can evaluate the ground
state spin correlation function. The correlation of matrix product
state can be easily derived by the transfer matrix technique, given
by,
\begin{equation}
\langle S^\mu_1S^\mu_r\rangle=(Tr
G^L)^{-1}Tr[Z(S^\mu)G^{r-2}Z(S^\mu)G^{L-r}],
\end{equation}
where $\mu=x,y,z$\cite{Tu2008}. Define $g=\sum_m\Gamma^m|m\rangle$,
then $G=g^\dagger\otimes g =\sum_{m}\Gamma^m\otimes \Gamma^m$, and
$Z(S^\mu)=g^\dagger\otimes S^\mu g$. As it's an isotropic magnet,
the correlation functions are the same in any directions. After some
detail calculation, we derive $\langle S^x_1S^x_r\rangle=\langle
S^y_1S^y_r\rangle=\langle S^z_1S^z_r\rangle=-20\times5^{-r}$ for
integer $r>1$. Therefore, the correlation length $\xi$ of the SZH
model equals to $1/\ln5$. This finite correlation length indicates
that the low-lying excitation in the SZH model is gapped, consistent
with the Haldane conjecture.

It is interesting to note that the correlation function of the SZH
model is negative-definite, with a correlation length
$\xi=1/ln5\sim0.61$. Consequently, the lattice constant is almost
twice of the correlation length, and the spins at neighboring sites
correlate extremely weakly. On the other hand, according to Arovas
\textit{et.al}'s work\cite{Arovas1988B}, the correlation length for
spin-2 AKLT is $1/\ln2$ which is roughly the lattice constant.
Therefore, although AKLT model is also a strongly disordered
antiferromagnet, the neighboring spins are closely correlated,
leading to the conventional staggering correlation function, say,
$\mathrm{S}_1\cdot\mathrm{S}_r\propto(-1)^r2^{-r}$.

Now we have two sets of models of 1-dimensional spin-2 chain with
exactly known Hamiltonians and ground state wavefunctions. The
differences between the AKLT and SZH model are not only the analytic
forms as they appear, but also the topological distinctions. The
same as topological insulator, the bulk topology is relevant to the
edge state of an open chain. For the spin-2 AKLT model, two solid
dots at each edge remain free. Symmetrical combination of these two
spin-1/2 dots results an edge spin with $S=1$. This boson-like edge
state is consistent with large-N theory of SU(N) quantum
antiferromagnets\cite{Ng1994}. However, the SZH model serves a
complement of the large-N analysis. For an open chain, the SZH
ground state is given by
\begin{equation}  \label{SZHopen}
|\Psi;i,j\rangle=\sum_{m_1,\ldots
m_N}(\Gamma^{m_1}\Gamma^{m_2}\ldots\Gamma^{m_N})_{ij}|m_1m_2\ldots
m_N\rangle.
\end{equation}
It explicitly shows that at each edge, there are four degrees of
freedom since the matrix product state is four dimensional.
Therefore, the edge state of SZH model is spin-3/2, \textit{ie.},
fermion-like. That is completely different from the AKLT model. As
the edge state is protected by topology, and is robust under
perturbation, the AKLT and SZH models belong to different
topological classes. It can be easily understood from the Berry
phase's language. Berry phase $\Phi_{BP}$ is the additional phase
when the spin winds around, which relates to the expectation value
of $T^2$ by $\exp(-i\Phi_{BP})=\langle T^2\rangle$, where $T$ is the
time reversal operator. It's well-known that $T^2=-1$ for half
integer spins such as $S=3/2$, while $T^2=1$ for integer spins such
as $S=1$. As a consequence, the Berry phases of the two models under
investigation differ by an angle of $\pi$. It is this difference
that makes topological distinction of AKLT and SZH models. From
another points of view, the edge spin determines the ground state
degeneracy (GSD)\cite{Wen1990a}, and naturally serves as a topology
index in analogy to fractional quantum Hall effect. These two models
thus have different topology index due to different edge spins.

Given the topological distinction of the two ground states, we
construct a model Hamiltonian interpolating between the AKLT and SZH
models:
\begin{equation}\label{Hamiltonian}
H(\alpha) =(1-\alpha )H^{AKLT}+\alpha H^{SZH}.
\end{equation}
Without loss of generality, we set $J_2=K_3=1$, $J_4=K_4=\beta$ in
the following. As the edge state is robust unless the gap closes,
there must exist one or several topological quantum phase
transitions (TQPT) where the gap closes and reopens in the evolution
of $\alpha$ from $0$ to $1$. This TQPT can be addressed by studying
the behavior of energy spectrum and correlation function at each
$\alpha$.

\begin{figure}[tbp]
\centerline{
    \includegraphics[height=4.0in,width=4.0in]{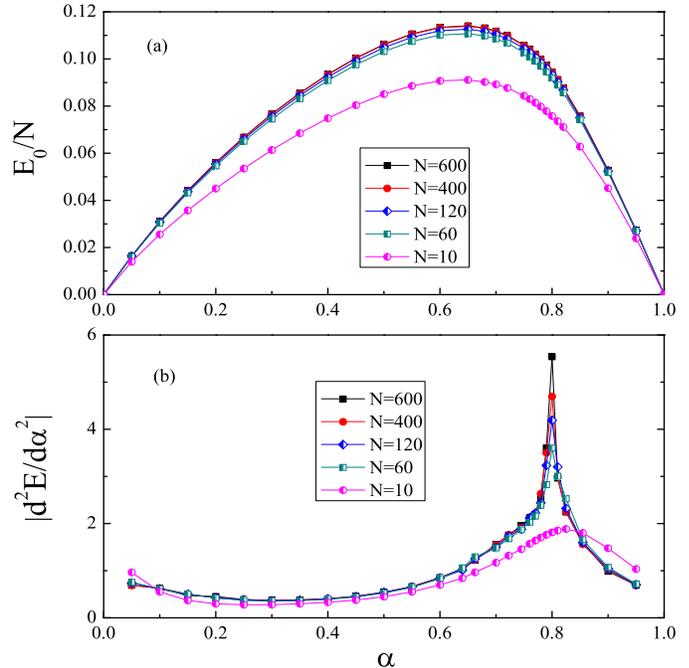}
    }
\caption{(color online) Ground state energy per site $E_0/N$ (a) and
the absolute value of its corresponding second derivative
$|d^2E/d\alpha^2|$ (b), as a function of $\alpha$, got by DMRG with
600 states and $\beta=1.0$ at different system sizes.
\label{Fig:Energy_Gradient}}
\end{figure}

\section{Numerical Results}
The {\it density matrix renormalization group} (DMRG) method is
employed\cite{White1992} in our study. For this purpose, it's
helpful to rewrite the projection operators explicitly in terms of
spin operators. Applying the identity
\begin{equation}
\bs_i\cdot\bs_j=\sum_{J=0}^{2S}[\frac{1}{2}J(J+1)-S(S+1)]P_J(ij),
\end{equation}
one can easily get
\begin{eqnarray}
P_2(ij)&=&\frac{1}{126}[-120\bss-14\bss^2+7\bss^3\nonumber\\
&&+\bss^4]\\
P_3(ij)&=&\frac{1}{360}[162\bss-7\bss^2-10\bss^3\nonumber\\
&&-\bss^4]+1\\
P_4(ij)&=&\frac{1}{2520}[90\bss+63\bss^2+14\bss^3\nonumber\\
&&+\bss^4]
\end{eqnarray}

For present study, we keep $m=600-1000$ states in the DMRG block
with more than $16$ sweeps to get a converged result, and the
truncation error is less than $10^{-7}$ in near the critical point,
and much less than $10^{-10}$ away from the critical point. We make
use of the open boundary condition (OBC) and the total number of
sites is $N=600$. To check the finite-size effect of the system, we
have studied the ground state energy per site $E_0/N$ and the
absolute value of the corresponding second derivative
$|d^2E/d\alpha^2|$ with respect to $\alpha$, with system size
$N=10-1000$ and $\beta=1$, as shown in
Fig.\ref{Fig:Energy_Gradient}. Such the ground state energy $E_0/N$
starts to converge, and the sharp peak of $|d^2E_0/d\alpha^2|$
appears at $N\geq 60$. Besides the ground state energy, we have also
calculated the correlation functions, and find that the finite-size
effect can be neglected at $N\geq 600$. Therefore, in the following
calculation, $N=600$ is adopted, for which the finite-size effect
can be completely neglected.


\begin{figure}[tbp]
\centerline{
    \includegraphics[height=4.2in,width=3.6in]{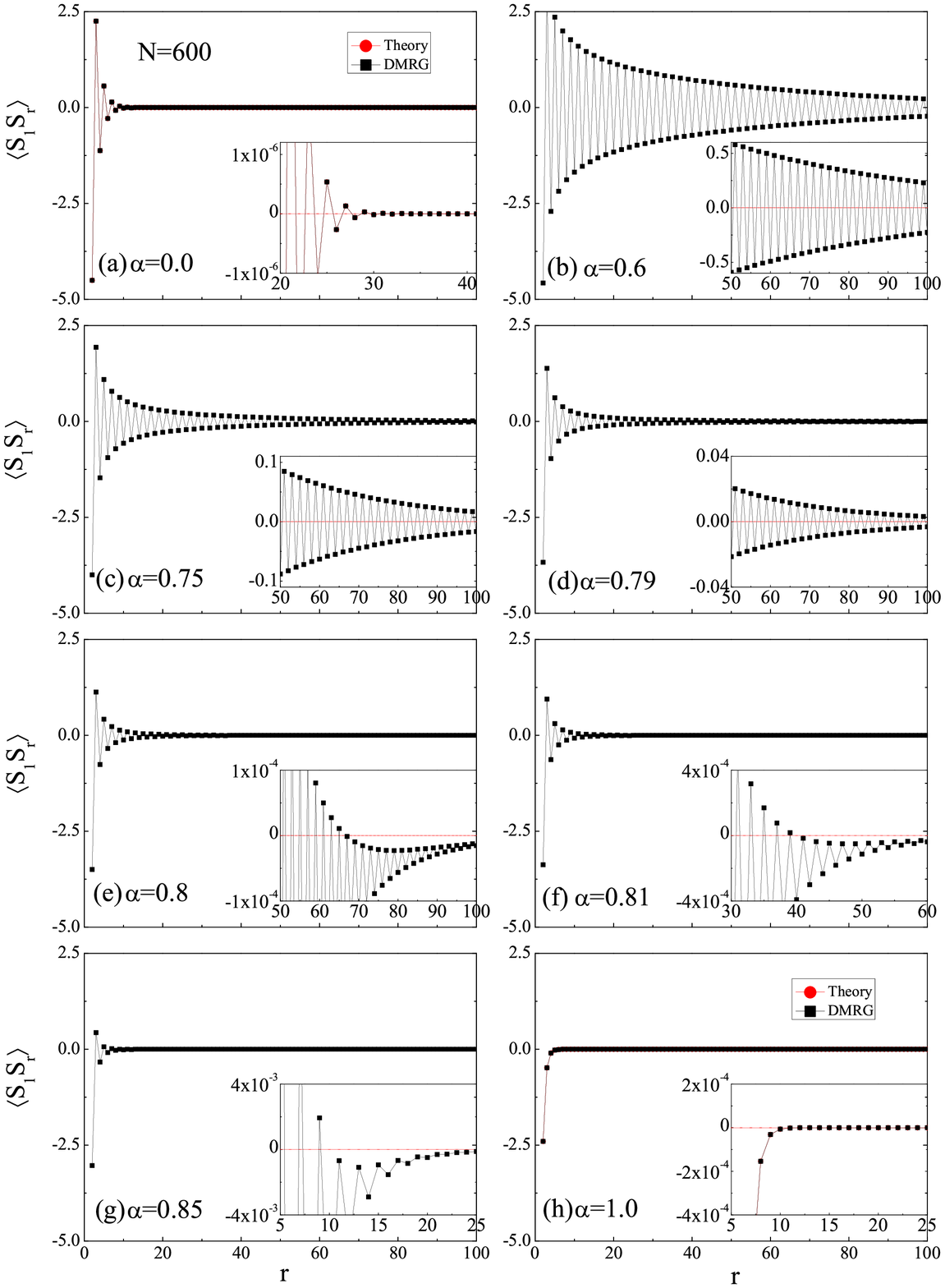}
    }
\caption{(color online) The real space spin-spin correlation
functions $\langle\textbf{S}_1\cdot \textbf{S}_r\rangle$ at
different $\alpha$, obtained by the DMRG with $N=600$ and
$\beta=1.0$. \label{Fig:SCor_N600}}
\end{figure}

Consistent with the discussion above, the existence of TQPT is
clearly supported by the numerical calculation, as shown in Fig.
\ref{Fig:Energy_Gradient}. The second derivative of the ground
energy has a sharp peak around $\alpha_c=0.80$, which indicates the
second order quantum phase transition addressed above. This is
further confirmed by the real-space spin-spin correlation function,
as shown in Fig.\ref{Fig:SCor_N600}. At $\alpha=0$ and $1$, the
correlation function is exactly the same with the analytical result.
When $\alpha<0.80$, the correlation behaves similarly with the AKLT
model, showing an oscillating behavior with respect the lattice
separation. However, when $\alpha>0.80$, the correlation behaves
similarly with the SZH model, showing an negative-definite behavior.
At the critical point, the correlation function undergoes an
qualitative change from AKLT to SZH.

\begin{figure}[tbp]
\centerline{
    \includegraphics[height=2.4in,width=3.6in]{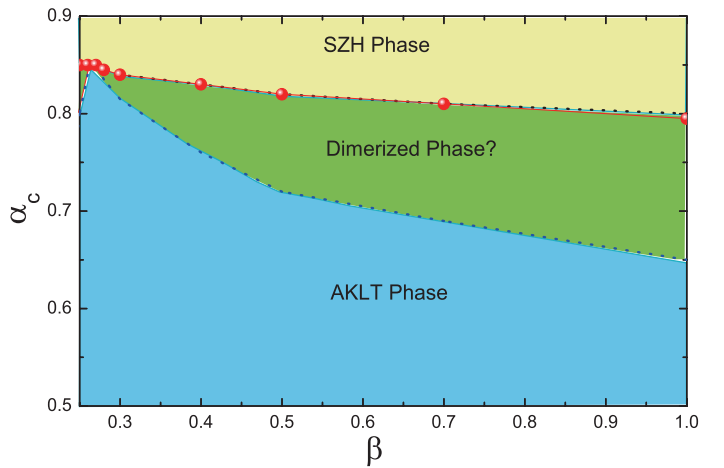}
    }
\caption{(color online) Ground state phase diagram of the model
Hamiltonian got by DMRG with $N=600$. The location of second order
phase transition for different $\beta$ is indicated by the red line.
The yellow region above red line belongs to SZH topological class,
while the light blue region belongs to AKLT class. Possible
dimerization is shown in the green region sandwiched between these
two phases. \label{Fig:dE_Dimer}}
\end{figure}

To get the ground state phase diagram, we also calculate the peak
position $\alpha_c$ of $|d^2E/d\alpha^2|$ as a function of $\beta$,
as shown in Fig.\ref{Fig:dE_Dimer}. Above the red line, the system
is topologically connected to SZH model, and therefore belongs to
the same topological class as SZH phase. One feature of this phase
is translational symmetric and has spin-$3/2$ excitations on the
edge. Similarly, the regions where $\alpha$ is small belong to the
same topological class as AKLT phase. Translational symmetry is also
respected, but with spin-$1$ excitations on the edge.

\begin{figure}[tbp]
\centerline{
    \includegraphics[height=3in,width=3.6in]{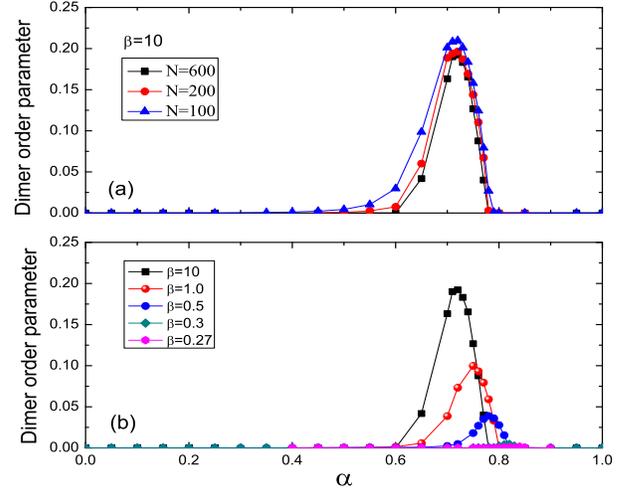}
    }
\caption{(color online) Dimer order parameters of the model
Hamiltonian with different choices of $\alpha$ and $\beta$.
(\textbf{a})The existence of dimerization is established by choosing
different system sizes $N$. (\textbf{b})Dimer order parameters for
different $\beta$. This parameter disappears for the critical
$\beta=0.27$.}\label{dimer}
\end{figure}

However, one thing we should keep in mind that even the model
Hamiltonian is translational invariant, spontaneous symmetry
breaking is also possible. One well known example is the
biliear-biquadratic model for spin-1 chain, whose Hamiltonian is
also written in terms of local projections, and translational
invariant. However, the emergence of the dimerized phase is
addressed in previous
works\cite{Klumper1990}\cite{Batchelor1990}\cite{Kennedy1992a}. In
order to quantitatively describe this issue, the following dimer
order parameter
\begin{equation}
D_\alpha=|\langle \mathbf{S}_i\mathbf{S}_{i+1}\rangle - \langle
\mathbf{S}_{i+1}\mathbf{S}_{i+2}\rangle|
\end{equation}
as a function of $\alpha$ is introduced, where $i$ labels the center
site in the spin chain so that one can minimize the possible
finite-size effect induced by open boundary condition. In the
calculation, the total site number $N$ is set to be an even number
to avoid potential ambiguity in the definition, in which case $i$ is
simply $N/2$. The indications of dimerization is plotted in
Fig.\ref{dimer}. It is shown explicitly that dimerization appears
with proper choices of parameters $\alpha$ and $\beta$. One can rule
out possible finite size effect as the dimer order parameter doesn't
scale with the system size $N$, shown in Fig.\ref{dimer}(a). When
$\beta$ decreases, the maximum amplitude of dimer order parameter
decreases as well, and the dimerization expands less and less
regions of $\alpha$ monotonously. At the critical point
$\beta=0.27$, dimer order parameter vanishes for any $\alpha$, and
the corresponding dimerization phase shrinks as well. The system
undergoes a quantum phase transition without spontaneous breaking of
translational symmetry. During this phase transition, the energy and
spin-spin correlation function have the same form as shown in Fig.
(\ref{Fig:Energy_Gradient}) and Fig. (\ref{Fig:SCor_N600}). However
it's worth mentioning that for Neel state of spin-2 chain, the
corresponding dimer order parameter would be 8. Therefore our dimer
order parameter is pretty small compared to strict antiferromagnets,
and the dimerization phase cannot be confirmed definitely. One
possibility is the gap between excitation state and ground state for
the present model is too small to be distinguished numerically. As a
result, certain dimerized excitation state enters into our results
and lead to such finite but tiny dimer order parameter. A promising
solution is to apply the periodic boundary condition (PBC) here, so
that one can rule out the possibility of dimerization acquired from
open boundary condition. Therefore, we have also done some
calculation by DMRG for system with PBC. For the system size
$N=100-200$ sites, we keep up to $m=3000$ states with truncation
error smaller than $10^{-8}$. Finally, we find that both OBC and PBC
systems give us consistent results.

Observed the possible presence of dimerization, one can readily work
out the phase diagram, see Fig.\ref{Fig:dE_Dimer}. The red curve
stands for critical points of $|d^2E/d\alpha^2|$, while the two
dashed dark curves are upper/lower bounds of dimerized phase of each
$\beta$ value. It shows the onset value for second order derivative
coincides with the upper bound of dimerized phase, so that they
describe the same phase transition between dimerized phase and SZH
phase. While on the AKLT side, the phase transition would be of
higher order that is invisible in Fig.\ref{Fig:Energy_Gradient}.
Despite the possible presence of dimerized phase, the system
undergoes a quantum phase transition without spontaneous breaking of
translational symmetry at $\alpha_c$ when $\beta=0.27$. As the
symmetry is unchanged in this case, this phase transition is
originated from the topology only.

In conclusion, the topological distinction of the AKLT model and SZH
model for the $S=2$ spin chain is presented in this work. A model
Hamiltonian as an interpolation between these two models is
introduced. The quantum phase transition of the model is protected
by topology, and established by DMRG calculation. The results
indicate the presence of a dimered phase between two topological
phases of AKLT and SZH. This dimered phase shrinks at critical value
of $\beta_c=0.27$. One would like to realize this topological phase
transition in real materials. On the AKLT side, we have a well
defined example already. $S=1$ edge spin is observed in $S=2$ chain
of CsCr$_{1-x}$Mg$_x$Cl$_3$\cite{Yamazaki1996a}. However, examples
on SZH side are still missing. Probably one can employ cold atom
techniques to realize TQPT proposed in this paper.

We thank Zheng-Cheng Gu, Hosho Katsura, Naoto Nagaosa, and Yong-Shi
Wu for insightful discussions. This work is supported by Ministry of
Education of China under the Grant No. B06011, the NSF of China, the
National Program for Basic Research of MOST-China, and SCZ is
supported by the NSF under grant numbers DMR-0904264. HCJ
acknowledges funding from Microsoft Station Q. In finishing this
work, we became aware of a parallel work\cite{Zheng2010a} which has
reached similar conclusion based on different approaches. They get a
brilliant result about the universality class of phase transition,
but the possible dimer order phase is not mentioned.

\end{document}